\documentclass[prc, twocolumn, reprint, superscriptaddress, showkeys]{revtex4-1}

\usepackage{latexsym}
\usepackage{graphicx}
\usepackage{amsmath}  
\usepackage{amssymb}
\usepackage{longtable}

\usepackage{natbib}

\begin{document}

\title{Determination of the Neutron-Capture Rate of $^{17}$C for the \textit{R}-process Nucleosynthesis}

\author{M.~Heine}
\email[E-mail:~]{m.heine@gsi.de}
\affiliation{Institut f\"ur Kernphysik, Technische Universit\"at Darmstadt, 64289 Darmstadt, Germany}

\author{S.~Typel}
\affiliation{GSI Helmholtzzentrum f\"ur Schwerionenforschung, 64291 Darmstadt, Germany} 

\author{M.-R.~Wu}
\affiliation{Institut f\"ur Kernphysik, Technische Universit\"at Darmstadt, 64289 Darmstadt, Germany} 

\author{T.~Adachi}
\affiliation{KVI-CART, University of Groningen, Zernikelaan 25, 9747 AA Groningen, The Netherlands}

\author{Y.~Aksyutina}
\affiliation{Institut f\"ur Kernphysik, Technische Universit\"at Darmstadt, 64289 Darmstadt, Germany}
\affiliation{GSI Helmholtzzentrum f\"ur Schwerionenforschung, 64291 Darmstadt, Germany}

\author{J.~Alcantara}
\affiliation{Dpt. de F\'{i}sica de Part\'{i}culas, Universidade de Santiago de Compostela, 15706 Santiago de Compostela, Spain}

\author{S.~Altstadt}
\affiliation{Goethe-Universit\"at Frankfurt am Main, 60438 Frankfurt am Main, Germany}

\author{H.~Alvarez-Pol}
\affiliation{Dpt. de F\'{i}sica de Part\'{i}culas, Universidade de Santiago de Compostela, 15706 Santiago de Compostela, Spain}

\author{N.~Ashwood}
\affiliation{School of Physics and Astronomy, University of Birmingham, Birmingham B15 2TT, United Kingdom}

\author{T.~Aumann}
\email[E-mail:~]{taumann@ikp.tu-darmstadt.de}
\affiliation{Institut f\"ur Kernphysik, Technische Universit\"at Darmstadt, 64289 Darmstadt, Germany}
\affiliation{GSI Helmholtzzentrum f\"ur Schwerionenforschung, 64291 Darmstadt, Germany}

% ...

\collaboration{R3B collaboration}
\noaffiliation

\begin{abstract}
With the R$^{3}$B-LAND setup at GSI we have measured exclusive relative-energy spectra of the Coulomb dissociation of $^{18}$C at a projectile energy around 425~AMeV on a lead target, which are needed to determine the radiative neutron-capture cross sections of $^{17}$C into the ground state of $^{18}$C. Those data have been used to constrain theoretical calculations for transitions populating excited states in $^{18}$C. This allowed to derive the astrophysical cross section $\sigma^{*}_{\mathrm{n}\gamma}$ accounting for the thermal population of $^{17}$C target states in astrophysical scenarios. The experimentally verified capture rate is significantly lower than those of previously obtained Hauser-Feshbach estimations at temperatures $T_{9}\leq{}1$~GK. Network simulations with updated neutron-capture rates and hydrodynamics according to the neutrino-driven wind model as well as the neutron-star merger scenario reveal no pronounced influence of neutron capture of $^{17}$C on the production of second- and third-peak elements in contrast to earlier sensitivity studies.
\end{abstract}


\maketitle

\section{Introduction}

Elements heavier than iron are mainly created in reactions in the slow (\textit{s-}) and rapid (\textit{r-}) neutron capture processes~\cite{kaeppeler2011, arnould2007} that are not suppressed by the Coulomb barrier at low energies \cite{burbidge1957}. The abundance pattern observed in ultra metal-poor stars~\cite{burris2000, sneden1996, westin2000, hill2002} attributed to the \textit{r-}process is remarkably close to solar in the range $56\leq{}Z\leq{}76$, which suggests a generic production mechanism in a unique astrophysical site.

This work addresses scenarios with nucleosynthesis flows sensitive to reaction rates of light neutron-rich nuclei~\cite{terasawa2001, sasaqui2005} found in core-collapse Type~II supernovae (SN) explosions of $M{}\leq{}2M_{\odot}$ progenitor stars in a rapid expansion scenario with a dynamical time scale $\tau_{\mathrm{dyn}}$ of a few milliseconds~\cite{sumiyoshi2000, otsuki2002}. A network study~\cite{terasawa2001} showed strong implications on \textit{r}-process nucleosynthesis as reactions involving light neutron-rich nuclei increase the efficiency for seed production and reduce the neutron-to-seed ratio drastically. The final heavy element abundances were found to change up to an order of magnitude as compared to calculations without light nuclai. The sensitivity to different reaction rates was investigated~\cite{sasaqui2005} and the neutron capture on $^{17}$C was considered critical as the rate was solely based on Hauser-Feshbach calculation. So far no experimental information on the neutron capture cross sections of $^{17}$C has been available.

Since $^{17}$C is unstable, the $^{17}\mathrm{C}(\mathrm{n},\,\gamma{})^{18}\mathrm{C}$ reaction is experimentally only accessible \textit{via} a time reversed reaction, \textit{e.g.}, by Coulomb excitation of $^{18}$C with subsequent neutron emission. The technique of Coulomb dissociation~\cite{baur1986} was established in reaction theory studies~\cite{shyam1991} while the accuracy for neutron-capture measurements as performed in this work was demonstrated~\cite{reifarth2008}.

In the current experiment electromagnetically induced transitions from the $^{18}$C ground state with spin-parity $J^{\pi}=0^{+}$ to all bound states in $^{17}$C with a neutron in the continuum were measured. This includes the first excited state $(J^{\pi}=1/2^{+})$ at 0.22~MeV excitation energy and the second excited state $(J^{\pi}=5/2^{+})$ at 0.33~MeV~\cite{ueno2013} besides the ground state $(J^{\pi}=3/2^{+})$. At the time being, Coulomb dissociation of an excited $^{18}$C beam cannot be measured. Hence, in the present analysis neutron capture from all bound states in $^{17}$C to the ground state in $^{18}$C was determined experimentally and complemented by theoretical calculations of transitions from all bound states in $^{17}$C to the first three excited states in $^{18}$C at 1.59~MeV $(J^{\pi}=2^{+})$, 2.50~MeV $(J^{\pi}=2^{+})$, and 3.99~MeV $(J^{\pi}=0^{+})$~\cite{stanoiu2004}.

The experimental setup is introduced in Section~\ref{sec:exp}. In Section~\ref{sec:res} exclusive energy-differential Coulomb dissociation data are presented and compared to theoretical calculations, that are described in detail in Section~\ref{sec:theo}. The calculation of reaction rates is delineated in Section~\ref{sec:ste} and the influence of the present results on nucleosynthesis simulations is discussed in Section~\ref{sub:imp}.

\section{Experiment}
\label{sec:exp}

\begin{figure}[htp]
  \center  
  \includegraphics[width=1.\textwidth]{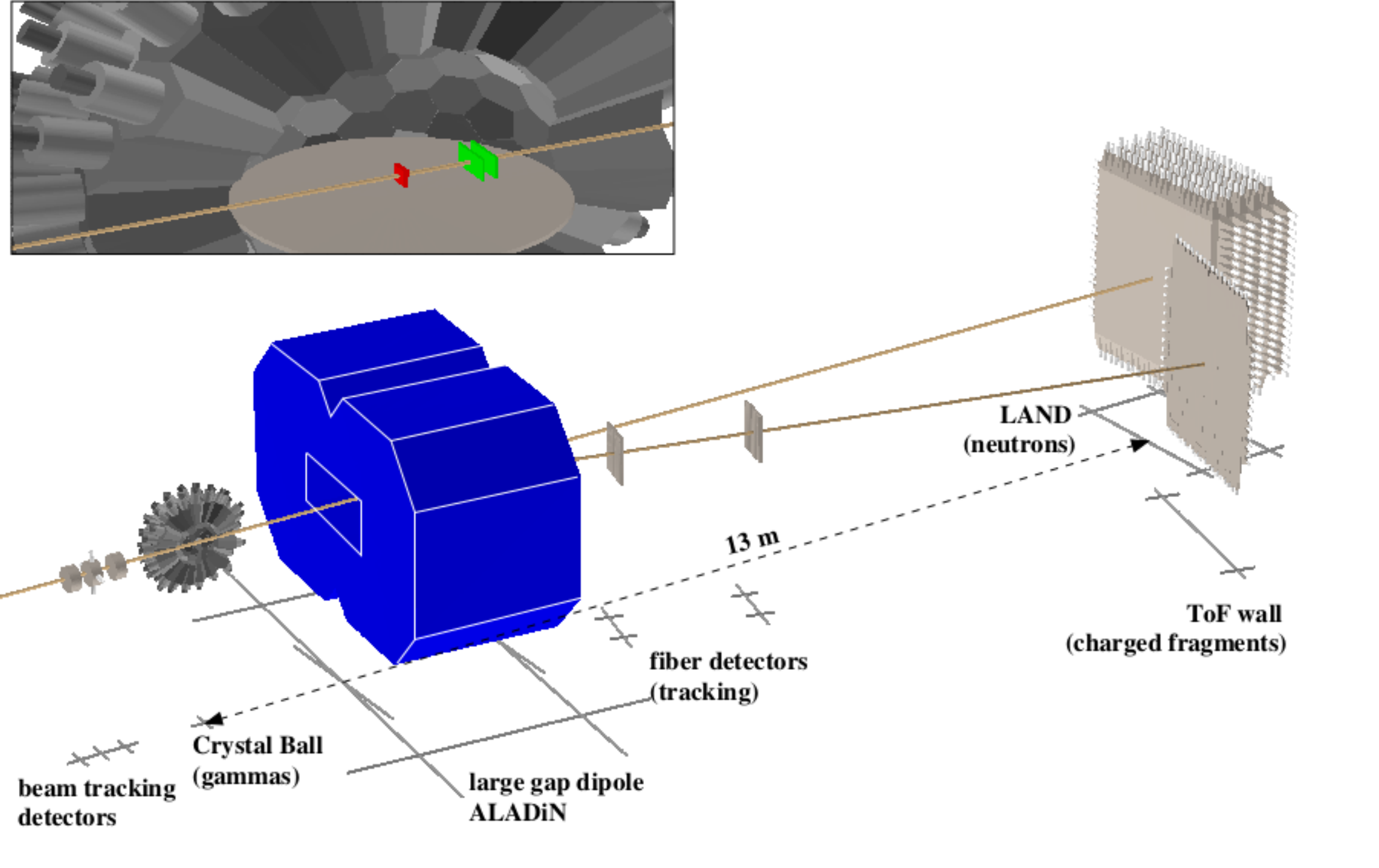} 
  \caption{\label{fig:s393}[color online] Relevant part of the R$^{3}$B-LAND setup for the present experiment with labels for the major components and indication of the particles to be detected. The inset shows the 3x3~cm$^{2}$ target (dark red) at the center of the gamma array, which is cut in half for illustration, and the fragment tracking detectors (light green) in front of the dipole magnet.}
\end{figure}

The measurements were carried out in complete inverse kinematics with the R$^{3}$B-LAND setup at GSI Helmholtzzentrum f\"u{}r Schwer\-ionen\-for\-schung GmbH. In Figure~\ref{fig:s393}, the relevant parts of the experimental setup are shown. It is designed for the coincident determination of the four-momenta of all reaction products from Time-of-Flight (ToF), position and energy-loss measurements. The separation of charged fragments is accomplished by A Large Acceptance Dipole magNet (ALADiN).

The $^{18}$C beam at 425~AMeV was produced by in-flight fragmentation of $^{40}$Ar at 490~AMeV in a 4~g/cm$^{2}$ thick beryllium target. Using the FRagment Separator (FRS)~\cite{geissel1992} ions with mass-to-charge ratio $(A/Z)$ of about three were selected and guided to the experimental setup. This secondary beam was identified event-by-event with respect to charge $Z$ and $A/Z$ using the ToF from the FRS to the experimental hall as well as by an energy-loss determination directly in front of the reaction target.

The $^{18}$C beam was directed onto the reaction target located at the center of the Crystal Ball array~\cite{habs1979} for $\gamma$-recognition indicating the population of excited states. Coulomb excitation reactions were induced in a 2145~mg/cm$^{2}$ lead target. In addition, data with a 935~mg/cm$^{2}$ carbon target and with an empty target frame were taken in order to account for background contributions from nuclear reactions in the target and non-specified interactions along the beam line, respectively.

Fragments were tracked \textit{via} position measurements with position sensitive silicon strip detectors~\cite{alcaraz2008}, fiber detectors~\cite{cub1998} and a ToF wall as well as by charge recognition and timing by the ToF wall. The LAND~\cite{blaich1992} was used for ToF and position measurements of neutrons.

\onecolumngrid

\twocolumngrid

\section{Analysis and Results}
\label{sec:res}

\begin{figure}[hpt]
  \center
  \includegraphics[width=1.\columnwidth]{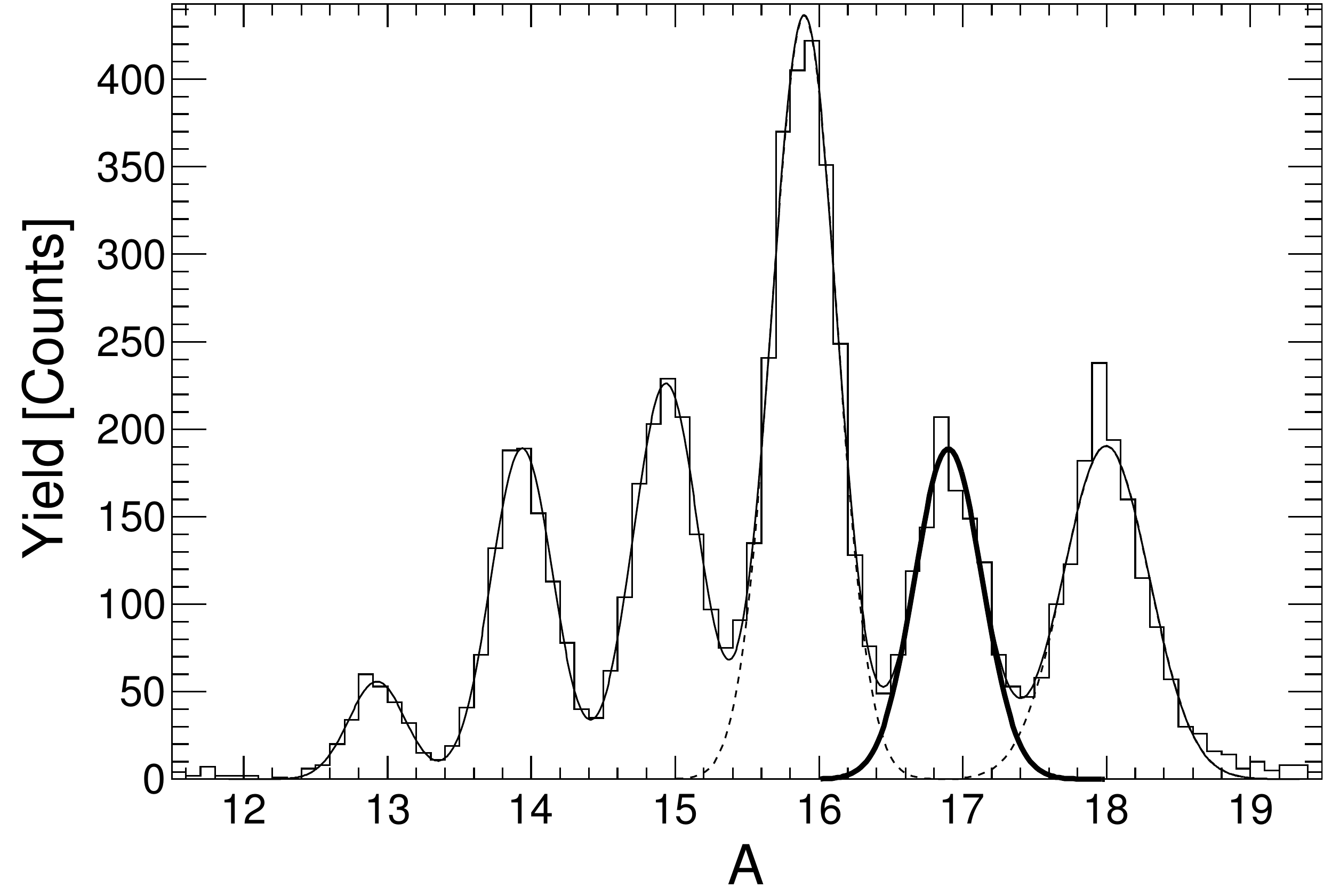}
  \caption{\label{fig:fra_A} Mass distribution of carbon fragments after $^{18}$C breakup on a lead target in coincidence with neutrons in LAND. The mass spectrum was fit by a multiple Gaussian and the one-neutron removal channel is indicated by the thick full line.}
\end{figure}

\noindent
Due to relativistic beam energies, the reaction products are extremely forward focused and moderately sized detectors are sufficient to perform 100$\%$~acceptance measurements. The reaction fragments were tracked through the $\vec{B}$-field of ALADiN by means of position measurements around the dipole as well as ToF and energy-loss measurements. In order to generate the fragment-mass spectrum the passage of the particles through the magnet was calculated accounting for the Lorentz force, while outside the magnetic field the projection of the flight path onto the tracking detectors was adjusted to experimental data by setting the fragment mass and velocity in an iterative procedure. In Figure~\ref{fig:fra_A}, the mass distribution of carbon fragments from $^{18}$C breakup on the lead target in coincidence with neutrons detected in LAND is displayed. Alongside with contributions mimicking non-reacted beam particles ($A=18$) several neutron removal channels are visible. One-neutron excitation reactions were selected around fragment mass $A=17$. The background around $A=18$ originates from break up reactions of $^{18}$C in detectors behind ALADiN and is removed in the analysis by a subtraction of data with the carbon target and without target.

\begin{figure}[hpt]
  \center
  \includegraphics[width=1.\columnwidth]{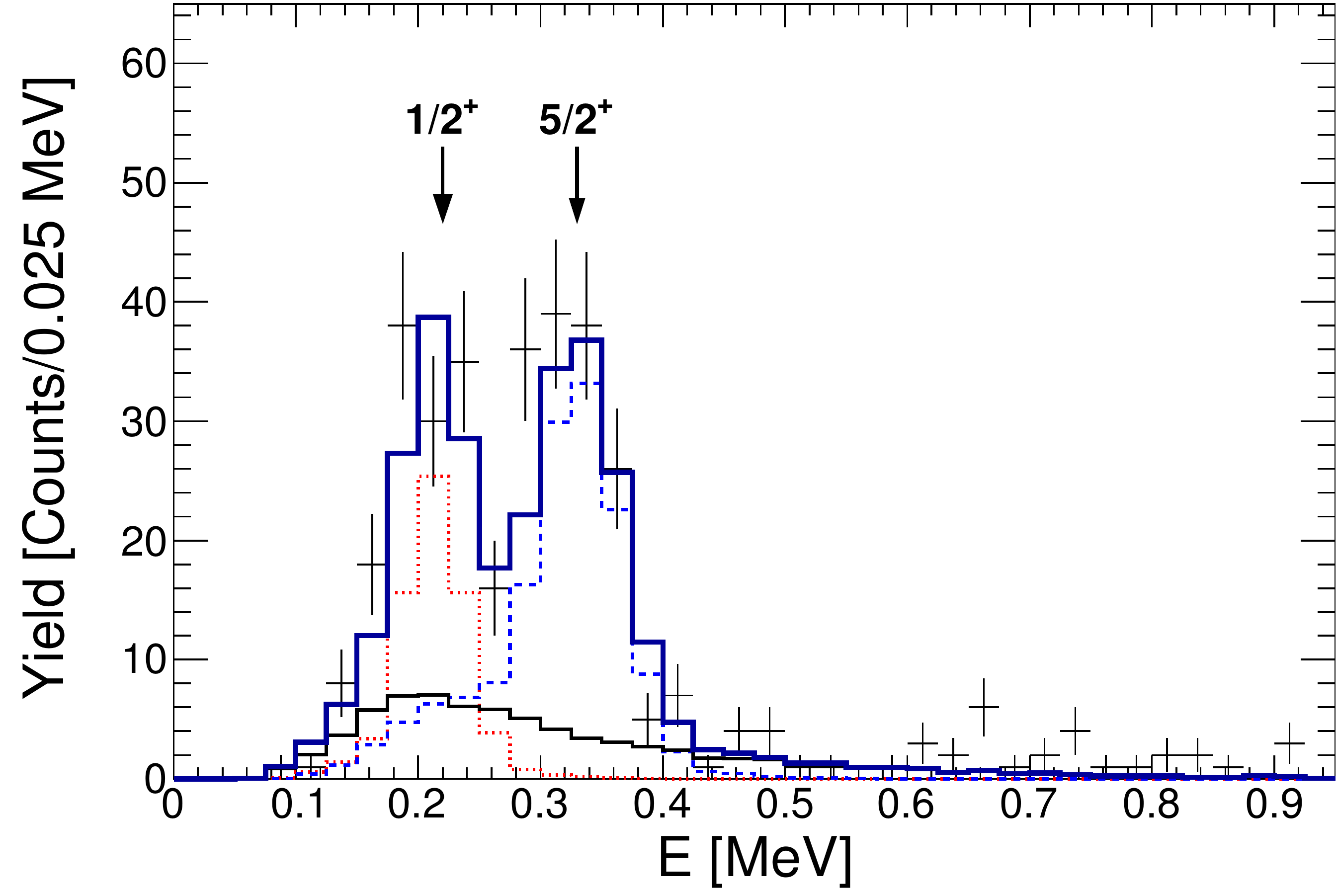}\llap{\makebox[3.cm][l]{\raisebox{1.95cm}[.1mm][.1mm]{\includegraphics[height=4cm]{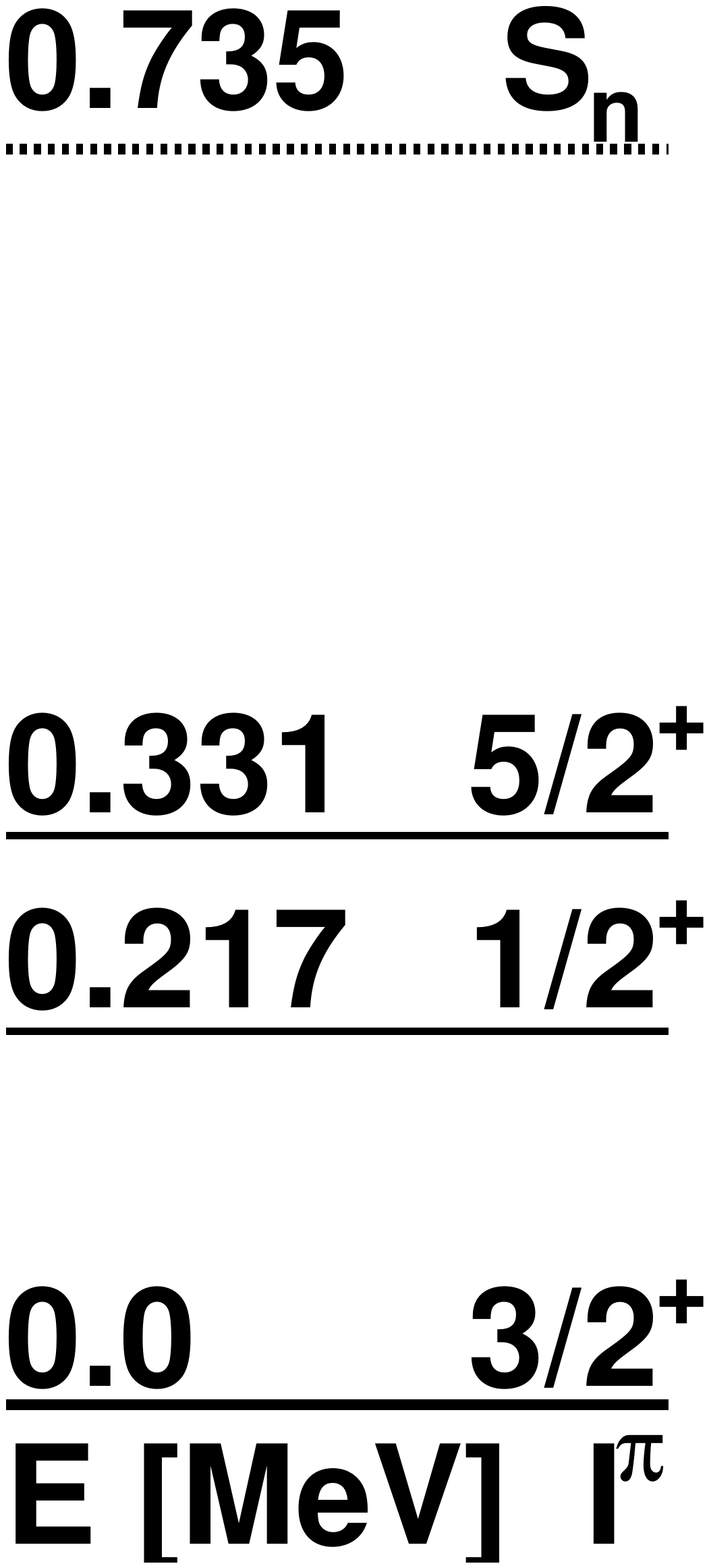}}}}
  \caption{\label{fig:xb}[color online] Identification of transitions to excited states in $^{17}$C from a comparison of experimental yields of lead target runs (crosses) with simulations (dotted red and dashed blue histograms for $1/2^{+}$ and $5/2^{+}$ states, respectively) and experimental atomic background (black histogram). The inset shows the level scheme of $^{17}$C~\cite{ueno2013}.}
\end{figure}

For identification of final excited states in $^{17}$C, the gamma detector response to emission of gammas from characteristic transitions in $^{17}$C was simulated and overlayed with experimental atomic background. The background was deduced from particles with mass $A=18$ in the fragment mass spectrum. The resulting response function was fit to experimental data as presented for $^{18}$C impinging on the lead target in Figure~\ref{fig:xb}.
The Doppler corrected gamma energy for clusters of crystals was summed up per event and for background reduction just emission in forward beam direction was analysed. The detection efficiency of gamma rays from the first excited state at 0.22~MeV and the second excited state at 0.33~MeV amounts to $39\%$ and $60\%$, respectively. The final states were identified as indicated in the figure. Misidentification due to the underlying atomic background from target de-excitation (black) is stronger for lower gamma energies and introduces a systematic uncertainty of 30$\%$.

Background from nuclear reactions in the target and non-specified interactions along the beam line was taken into account by subtracting normalized data with carbon target and without target inserted, respectively, from lead target runs. To correct for the contribution of nuclear reactions in these runs, exclusive experimental nuclear reaction channels, in the presented case reactions with proton removal, were utilized. These final states cannot be populated by electromagnetic excitation at present beam energy. Carbon target data were scaled to compensate these channels in lead target data. This experimental scaling factor $\alpha_{Pb}=1.65$ compares to the empirical value $\alpha_{Pb}=1.38$ derived from the ``factorisation model''~\cite{boretzky2003} with non-specified uncertainties from the systematics, and $\alpha_{Pb}=1.74$ derived from the black-disk model for peripheral reactions simply using nuclear radii for scaling. The value of $\alpha_{Pb}=1.65$ derived in this experiment was preferred in the analysis.

The experimental one-neutron evaporation cross sections are listed in Table~\ref{tab:xs_c2s}, where the total electromagnetic cross section was obtained from the normalized number of $^{17}$C fragments, accounting for the background mentioned before. Ground state transitions were derived after subtracting efficiency corrected contributions from excited states identified in the gamma spectrum. The final states in $^{17}$C are fairly uniformly populated from Coulomb dissociation. 

The experimental spectroscopic factor $S_{i}(J^{\pi}_{\rm core},l_{j})$ of transition $i$ including $^{17}$C core states with total angular momentum and parity $J_{core}^{\pi}$ and angular momentum of the valence neutron $l_{j}$ was obtained by dividing the measured partial cross section $\sigma_{\mathrm{exp}}$ by the corresponding theoretical single particle cross section $\sigma_{\mathrm{sp}}$ (see Section~\ref{sec:theo})
\begin{align}
 S_{i}(J^{\pi}_{\rm core},l_{j}) & = \frac{\sigma_{\mathrm{exp}}}{\sigma_{\mathrm{sp}}}.
\end{align}
The spectroscopic factors in Table~\ref{tab:xs_c2s} are consistent with results from a knockout experiment~\cite{kondo2009} and shell-model calculations~\cite{brown1986} within a $3\sigma$-range. The quantitative effect of using experimental instead of shell model amplitudes for the final neutron-capture rates is discussed in Section~\ref{sec:ste}.

\section{Theoretical Calculations}
\label{sec:theo}

In the calculation of exclusive energy-differential single particle cross sections and for comparison with experimental data, photo-absorption cross sections $\sigma^{\mathrm{photo}}_{E1}$ were derived with the CDXS+ code~\cite{typel2007} and converted subsequently into Coulomb dissociation cross sections based on equivalent photon theory~\cite{baur1986} as
\begin{align}
  \frac{\mathrm{d}\sigma_{CD}}{\mathrm{d}E_{\gamma}} & = \frac{n_{E1}}{E_{\gamma}} \cdot \sigma^{\mathrm{photo}}_{E1} \label{eq:cd_pa} \: .
\end{align}
Here, $E_{\gamma}$ denotes the excitation energy and $n_{E1}$ the number of virtual $E1$ photons.

For the calculation of single-particle cross sections, the ${}^{18}$C nucleus is described in a simple potential model assuming a core valence-neutron picture. Choosing the lowest states in ${}^{17}$C as possible core states, there are various ways to couple the spins of the core and the valence neutron in the \textit{sd}-shell to the total angular momentum and parity of ${}^{18}$C. For the three lowest states in ${}^{17}$C we assumed total angular momenta and parities of $3/2^{+}$, $1/2^{+}$, and $5/2^{+}$ with excitation energies of $0.00$~MeV, $0.22$~MeV and $0.33$~MeV, respectively~\cite{ueno2013}. The spin assignments are in line with those derived in the analysis of transverse-momentum distributions of excited states~\cite{kondo2009}. Furthermore, we considered the four lowest states ($i=0,1,2,3$) in ${}^{18}$C~\cite{stanoiu2004}: the ground state ($J^{\pi}_{0}=0^{+}$), the first excited state ($J^{\pi}_{1}=2^{+}$) at 1.59~MeV, the second excited state ($J^{\pi}_{2}=2^{+}$) at 2.50~MeV, and the third excited state ($J^{\pi}_{3}=0^{+}$) at an excitation energy of 3.99~MeV. The two $0^{+}$ states ($i=0,3$) can be decomposed as a linear combination

\onecolumngrid

\begin{center}
  \begin{table}[hpt]
    \caption{\label{tab:xs_c2s}Experimental cross sections ($\sigma_{\mathrm{exp}}$) and single-particle Coulomb excitation cross sections ($\sigma_{\mathrm{sp}}$) for $^{17}$C core states with total angular momentum and parity $J_{core}^{\pi}$ and angular momentum of the neutron $l_{j}$ calculated with the CDXS+ code~\cite{typel2007} in plane-wave approximation. The $S_{i}(J^{\pi}_{\rm core}, l_{j})$ are spectroscopic factors for one-neutron removal in $^{18}$C. The experimental $S_{i}(J^{\pi}_{\rm core}, l_{j})$ from Coulomb dissociation are compared to results taken from a knockout experiment~\cite{kondo2009} and shell model calculations~\cite{brown1986} in the $psd$ model space with the WBP interaction.}
    \centering
    \begin{tabular}{c c c c@{$\,\pm\,$}r@{$\,\pm\,$}r c c@{$\,\pm\,$}r@{$\,\pm\,$}r c r@{$\,\pm\,$}l c c}
      \hline\hline
      $E$~[MeV] & $J_{core}^{\pi}$ & $l_{j}$ & \multicolumn{3}{c}{$\sigma_{\mathrm{exp}}$ [mb]} & $\sigma_{\mathrm{sp}}$ [mb] & \multicolumn{8}{c}{$S_{i}(J^{\pi}_{\rm core}, l_{j})$} \\
      \hline
      & & & \multicolumn{3}{r}{} & & \multicolumn{3}{c}{Coulomb}  & & \multicolumn{2}{c}{Knockout} & & Shell Model \\
      \cline{4-6}
      \cline{8-10}
      \cline{12-13}
      \cline{15-15}
      & & & \multicolumn{3}{r}{stat. sys.} & & \multicolumn{3}{r}{stat. $\;\,$ sys.}  & & \multicolumn{2}{r}{} & & \\
      & & & \multicolumn{3}{c}{} & & \multicolumn{3}{c}{}  & & \multicolumn{2}{c}{} & & \\
      0.0  & $3/2^{+}$ & 2 & 32 & $\;\;$13 & $\;\,$5 & 27 & 1.18 & 0.48 & 0.19 & & \multicolumn{2}{c}{$\quad\leq 0.67$} & & 0.10 \\
      0.22 & $1/2^{+}$ & 0 & 40 & 8  & 5 & 75 & 0.52 & 0.11 & 0.07  & & 0.39 & 0.07 & & 0.65 \\
      0.33 & $5/2^{+}$ & 2 & 43 & 6  & 1 & 25 & 1.74 & 0.24 & 0.04 & & 2.39 & 0.27 & & 2.80 \\
      & & & \multicolumn{3}{c}{} & & \multicolumn{3}{c}{} & & \multicolumn{2}{c}{} & & \\
      total  & &   &          115 &  \multicolumn{2}{l}{$\;\,$8}  & & \multicolumn{3}{c}{} & & \multicolumn{2}{c}{} & & \\
      \hline\hline
    \end{tabular}
  \end{table}
\end{center}

\twocolumngrid

\begin{eqnarray} \label{equ:sd_sp_0}
  |^{18}\mathrm{C}(0^{+})\rangle \, & & = A_{i}(3/2^{+},d_{3/2})  |^{17\!}\mathrm{C}(3/2^{+})\otimes\nu 0d_{3/2}\rangle \\
  \nonumber & & + \,A_{i}(1/2^{+},s_{1/2})  |^{17\!}\mathrm{C}(1/2^{+})\otimes\nu 1s_{1/2}\rangle \\
  \nonumber & & + \, A_{i}(5/2^{+},d_{5/2})  |^{17\!}\mathrm{C}(5/2^{+})\otimes\nu 0d_{5/2}\rangle \: ,
\end{eqnarray}
\noindent
with spectroscopic amplitudes $A_{i}(J^{\pi}_{\rm core},l_{j})$ that are linked to spectroscopic factors as $S_{i}(J^{\pi}_{\rm core},l_{j}) = [A_{i}(J^{\pi}_{\rm core},l_{j})]^{2}$. For the two excited $2^{+}$ states ($i=1,2$), the decomposition is more complicated with eight contributions
\begin{eqnarray} \label{equ:sd_sp_2}
  |^{18}\mathrm{C}(2^{+})\rangle \, & & = A_{i}(3/2^{+},d_{3/2})  |^{17\!}\mathrm{C}(3/2^{+})\otimes\nu 0d_{3/2}\rangle \\
  \nonumber & & + \,A_{i}(3/2^{+},d_{5/2})  |^{17\!}\mathrm{C}(3/2^{+})\otimes\nu 0d_{5/2}\rangle \\
  \nonumber & & + \,A_{i}(3/2^{+},s_{1/2})  |^{17\!}\mathrm{C}(3/2^{+})\otimes\nu 1s_{1/2}\rangle \\
  \nonumber & & + \,A_{i}(1/2^{+},d_{3/2})  |^{17\!}\mathrm{C}(1/2^{+})\otimes\nu 0d_{3/2}\rangle \\
  \nonumber & & + \,A_{i}(1/2^{+},d_{5/2})  |^{17\!}\mathrm{C}(1/2^{+})\otimes\nu 0d_{5/2}\rangle \\
  \nonumber & & + \,A_{i}(5/2^{+},d_{3/2})  |^{17\!}\mathrm{C}(5/2^{+})\otimes\nu 0d_{3/2}\rangle \\
  \nonumber & & + \,A_{i}(5/2^{+},d_{5/2})  |^{17\!}\mathrm{C}(5/2^{+})\otimes\nu 0d_{5/2}\rangle \\
  \nonumber & & + \,A_{i}(5/2^{+},s_{1/2})  |^{17\!}\mathrm{C}(5/2^{+})\otimes\nu 1s_{1/2}\rangle \: .
\end{eqnarray}

The wave function of the valence neutron in each component was determined by solving the Schr\"{o}dinger equation for the neutron-core relative motion using a Woods-Saxon potential with radius $r=1.25\cdot A^{1/3}$~fm ($A=18$) and diffuseness parameter $a=0.65$~fm. The corresponding potential depth $V(J^{\pi}_{\rm core},l_j)$ was adjusted to reproduce the experimental neutron separation energies taking the excitation energy of the core into account. Explicit values are given in Table~\ref{tab:pot_depths}. The spectroscopic amplitudes $A_{i}(J^{\pi}_{\rm core},l_{j})$ were calculated using the shell model code \texttt{OXBASH}~\cite{brown1986} in the $psd$ model space with the WBP interaction~\cite{warburton1992}.

\begin{table}
  \centering
  \caption{\label{tab:pot_depths} Depths of the Woods-Saxon potentials in the calculation of the neutron wave function for the different components of the ground and excited states in ${}^{18}$C, see text for details.}
  \begin{tabular}{cccccc}
    \hline \hline
    $i$ & $J^{\pi}_{i}$ & $J^{\pi}_{\rm core}$ & $l_{j}$ & $S_{i}(J^{\pi}_{\rm core}, l_{j})$ & $V(J^{\pi}_{\rm core},l_{j})$ [MeV]\\
    \hline
    0          & $0^{+}$  & $3/2^{+}$ & $d_{3/2}$ & 0.10 & 53.504156 \\
               &          & $1/2^{+}$ & $s_{1/2}$ & 0.65 & 52.834161 \\
               &          & $5/2^{+}$ & $d_{5/2}$ & 2.80 & 54.157566 \\
    \hline
     1         & $2^{+}$  & $3/2^{+}$ & $d_{3/2}$ & 0.01 & 50.246199 \\
               &          & $3/2^{+}$ & $d_{5/2}$ & 1.08 & 50.246199 \\
               &          & $3/2^{+}$ & $s_{1/2}$ & 0.02 & 48.155175 \\
               &          & $1/2^{+}$ & $d_{3/2}$ & 0.08 & 50.690976 \\
               &          & $1/2^{+}$ & $d_{5/2}$ & 0.17 & 50.690976 \\
               &          & $5/2^{+}$ & $d_{3/2}$ & 0.08 & 50.943124 \\
               &          & $5/2^{+}$ & $d_{5/2}$ & 0.44 & 50.943124 \\
               &          & $5/2^{+}$ & $s_{1/2}$ & 0.23 & 49.078979 \\
    \hline
     2         & $2^{+}$  & $3/2^{+}$ & $d_{3/2}$ & 0.09 & 48.239374 \\
               &          & $3/2^{+}$ & $d_{5/2}$ & 0.13 & 48.239374 \\
               &          & $3/2^{+}$ & $s_{1/2}$ & 0.53 & 45.339128 \\
               &          & $1/2^{+}$ & $d_{3/2}$ & 0.01 & 48.707575 \\
               &          & $1/2^{+}$ & $d_{5/2}$ & 0.07 & 48.707575 \\
               &          & $5/2^{+}$ & $d_{3/2}$ & 0.02 & 48.972373 \\
               &          & $5/2^{+}$ & $d_{5/2}$ & 0.07 & 48.972373 \\
               &          & $5/2^{+}$ & $s_{1/2}$ & 0.04 & 46.400410 \\
    \hline
     3         & $0^{+}$  & $3/2^{+}$ & $d_{3/2}$ & 0.05 & 44.688354 \\
               &          & $1/2^{+}$ & $s_{1/2}$ & 1.20 & 40.065049 \\
               &          & $5/2^{+}$ & $d_{5/2}$ & 0.15 & 45.523757 \\
     \hline \hline
  \end{tabular}
\end{table}

In the evaluation of electromagnetic transitions from ${}^{18}$C bound states to ${}^{17}$C+neutron continuum states and \textit{vice versa}, only the \textit{E1} multipolarity was considered because electric transitions with higher multipolarities are strongly suppressed due to the smaller effective charges~\cite{bertulani1988b}. Thus only negative-parity states are relevant in the continuum. All possible couplings of a ${}^{17}$C core state with a neutron in \textit{p}- or \textit{f}-waves were taken into account. The scattering wave functions were calculated without a neutron-core interaction corresponding to a plane-wave approximation. In principle, a finite strength of the interaction in these channels can be expected but without precise experimental information on resonant states, it cannot be determined unambiguously. When potential depths of similar size as for the bound states are chosen, the appearance of arbitrary resonant states cannot be excluded. They would strongly distort the theoretical Coulomb breakup spectrum but there are no hints in this direction from the present experiment. This is in line with no resonances expected for approximately 700~keV above the neutron separation threshold  in $^{18}$C from theoretical calculations~\cite{stanoiu2004, herndl1999}. Coulomb breakup cross sections were calculated in the semiclassical approach using the relativistic straight-line approximation that is valid at high beam energies.

\begin{figure}[hpt]
  \center
  \flushleft{a)} \\
  \includegraphics[width=1.\columnwidth]{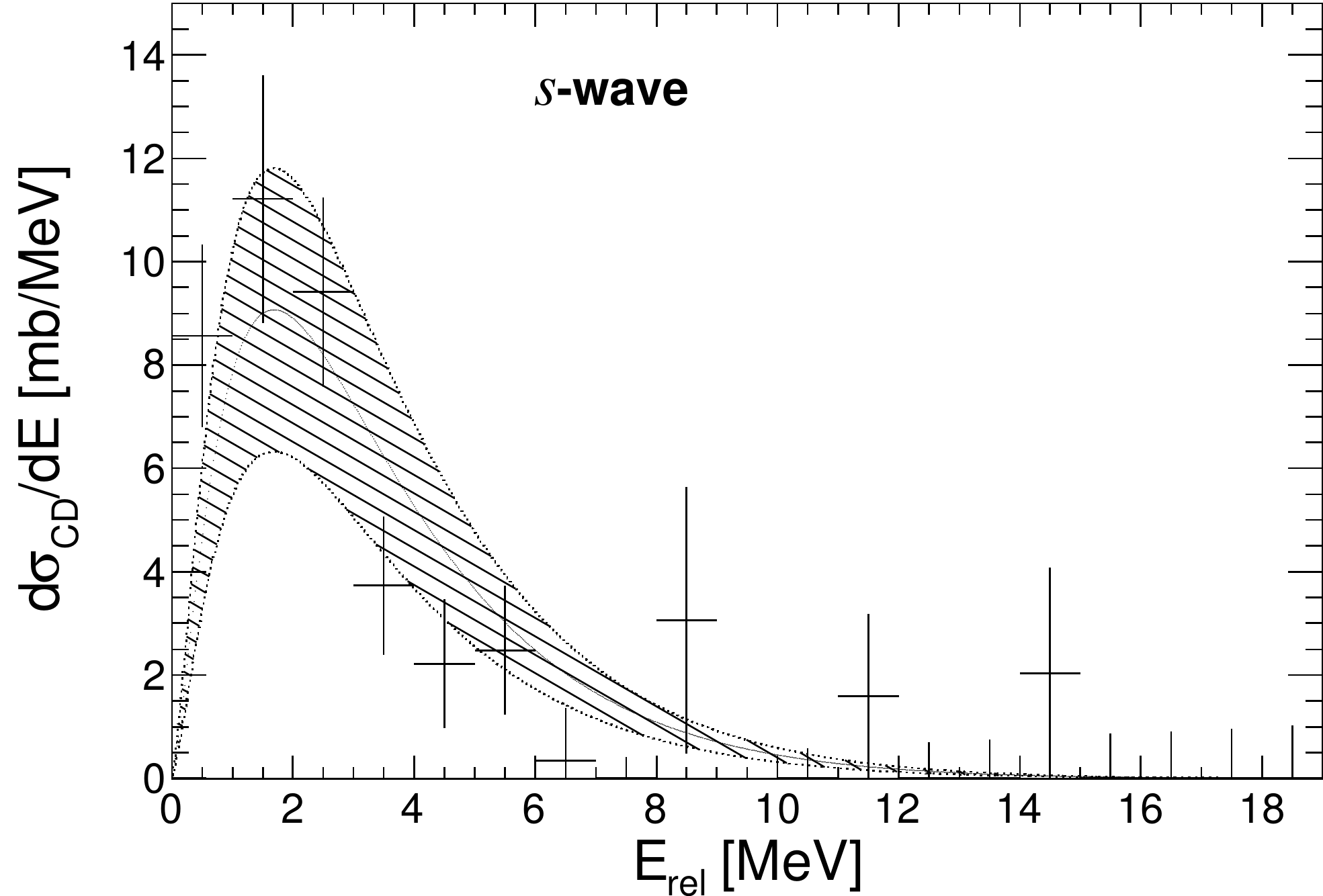}
  \vspace*{-.6cm}
  \flushleft{b)} \\
  \includegraphics[width=1.\columnwidth]{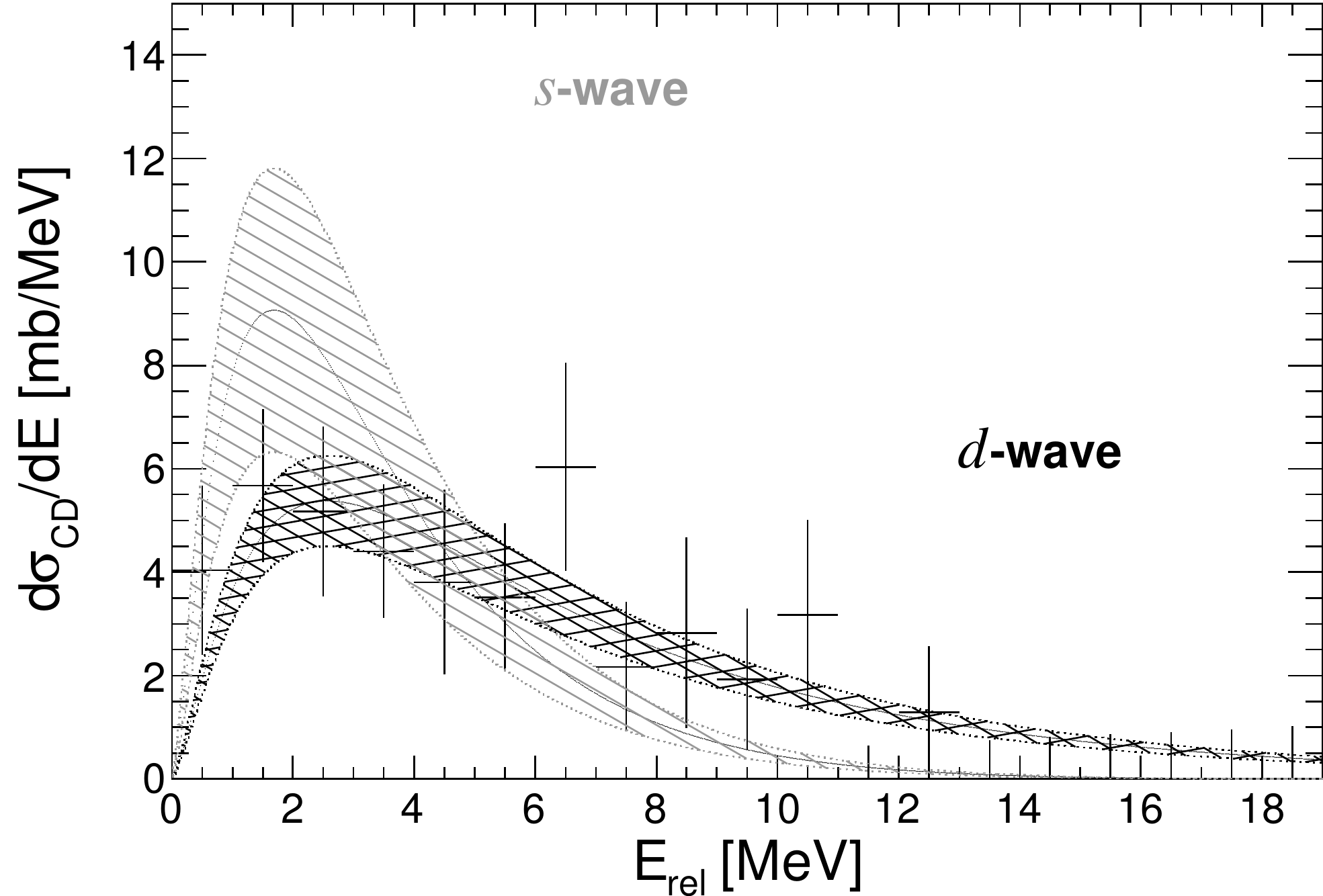}
  \caption{\label{fig:s_de} Experimental exclusive differential Coulomb dissociation cross sections (crosses) of a) \textit{s}-wave excitation populating the first excited state $(1/2^{+})$ in $^{17}$C and b) \textit{d}-wave excitation in $^{18}$C populating the second excited state $(5/2^{+})$ in $^{17}$C with respect to the ${}^{17}$C-n relative energy. For comparison also the \textit{s}-wave excitation is shown in the lower panel. Ambiguities in the identification of excited states in $^{17}$C with the gamma detector are indicated by the width of the bands of the semiclassical model calculations.}
\end{figure}

The Coulomb-dissociation calculations are compared to background-subtracted experimental exclusive differential cross sections with respect to the relative energy for the case of transitions to the $1/2^{+}$ state in $^{17}$C at 0.22~MeV in Figure~\ref{fig:s_de} in the upper panel. Data show a typical behavior of non-resonant excitation to \textit{s}-wave states in the continuum. This is similar to observations in the Coulomb breakup of neutron halo nuclei~\cite{aumann2013} with a cross section maximum at low relative energy. Broader \textit{d}-wave distributions were obtained for transitions attributed to the $3/2^{+}$ ground state as well as the $5/2^{+}$ excited state. The latter is shown in the lower panel in Figure~\ref{fig:s_de}. The calculations were scaled to the integral experimental Coulomb breakup cross sections and the experimental spectroscopic strength listed in Table~\ref{tab:xs_c2s}.

\section{Thermonuclear Reaction Rate}
\label{sec:ste}

The photo absorption cross sections $\sigma^{\mathrm{photo}}_{E1}$ were converted into neutron capture cross sections $\sigma_{\mathrm{n}\gamma}$ with the detailed balance theorem~\cite{baur1986} 
\begin{align}
  \sigma_{\mathrm{n}\gamma}(E_{\rm rel}) & = \frac{2(2J_{^{18}\mathrm{C}}^{\pi}+1)}{(2J_{^{17}\mathrm{C}}^{\pi}+1)(2J_{\mathrm{n}}^{\pi}+1)} \frac{k^{2}_{\gamma}}{k^{2}_{\rm rel}} \cdot \sigma^{\mathrm{photo}}_{E1}, \label{eq:ng_pa}
\end{align}
exploiting that the modulus squares of the matrix elements for exclusive transitions are the same in time-reversed processes. Here $J^{\pi}_{i}$ are the spins of the parti\-ci\-pating nuclei, $k_{\gamma}$ and $k_{\rm rel}$ are the momentum of the $E1$ photon and the momentum of relative motion in the core valence-neutron system, respectively.

\begin{figure}[hbt]
  \center
  \includegraphics[width=1.\columnwidth]{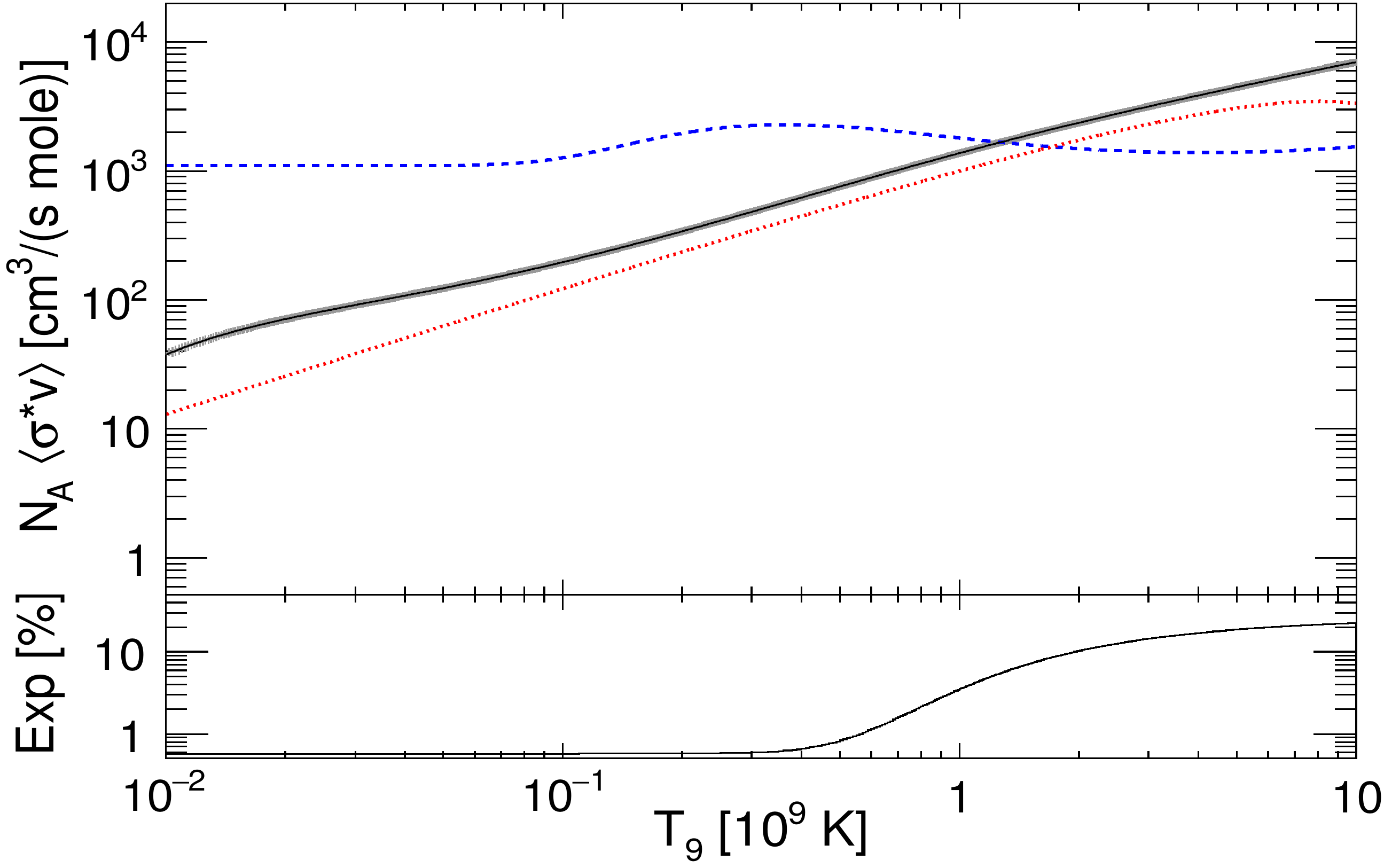}
  \caption{\label{fig:rr} [color online] (Upper panel) Reaction rate for neutron capture on ${}^{17}$C with respect to the stellar temperature $T_{9}$. Present data (grey band) are compared to Hauser-Feshbach rates~\cite{sasaqui2005} (dashed blue line) and a direct capture model~\cite{herndl1999} (dotted red line) calculation. In the lower panel the actual contribution of experimental data, \textit{i.e.} transitions to the ground state in $^{18}$C, is displayed.}
\end{figure}

In astrophysical applications, the thermonuclear reaction rate is needed accounting for the velocity distribution of the neutrons in thermal equilibrium with the stellar environment. It is given in cm$^{3}$s$^{-1}$ by
\begin{align}
  \langle \sigma_{\mathrm{n}\gamma} v \rangle & = \sqrt{\frac{8}{\pi \mu (k_{\mathrm{B}}T)^{3}}} \int^{\infty}_{0} \mathrm{d}E \sigma_{\mathrm{n}\gamma}(E)\,E\,\mathrm{e}^{-\frac{E}{k_{\mathrm{B}}T}} \label{eq:sRR_t} \: .
\end{align}
\noindent
Here, $k_{\mathrm{B}}$ is the Boltzmann constant, $\mu$ is the reduced mass of the core valence-neutron system and $T$ is the temperature of the gas at the astrophysical site. Considering the thermal population of the core states \textit{j} in $^{17}$C with excitation energy $E_{j}$ and spin $J_{j}$ the neutron capture cross section $\sigma_{\mathrm{n}\gamma}$ for a single transition in equation (\ref{eq:sRR_t}) has to replaced by the averaged astrophysical cross section of non-resonant transitions as~\cite{rauscher2000}
\begin{align}
  \sigma_{\mathrm{n}\gamma}^{*}(E) & = \frac{\sum_{j}(2J_{j} + 1) \exp(-E_{j}/k_{\mathrm{B}}T)\cdot \sum_{i} \sigma_{\mathrm{n}\gamma}^{ij} (E)}{\sum_{j}(2J_{j} + 1) \exp(-E_{j}/k_{\mathrm{B}}T)} \label{eq:sig_ste} \: .
\end{align}
\noindent
Here, the $\sigma_{\mathrm{n}\gamma}^{ij}$ are the neutron capture cross sections of particular target states \textit{j} in $^{17}$C to all considered states \textit{i} in $^{18}$C. In the present analysis all bound states in $^{17}$C and $^{18}$C were taken into account (see Table~\ref{tab:pot_depths}). Note that the partition function (the denominator) in Equation~(\ref{eq:sig_ste}) at $T_{9}=T/(1GK)=1$ is 1.072 when normalized to the ground state, \textit{i.e.} it yields a thermal population of 93\% of $^{17}$C in the ground state. The stellar reaction rate $\langle \sigma_{\mathrm{n}\gamma}^{*} v \rangle$ for neutron capture on $^{17}$C in cm$^{3}$/(mole$\cdot$s) is shown in Figure~\ref{fig:rr} as a function of the stellar temperature $T_{9}$. The upper panel in Figure~\ref{fig:rr} displays the present capture rate (grey band) in comparison to a parametrization from a Hauser-Feshbach estimation~\cite{sasaqui2005} (blue dashed curve) and a parametrization using a direct capture model~\cite{herndl1999} (red dotted curve). The present data set is approximately proportional to $T$ that is characteristic for negative parity-state capture~\cite{fowler1967}, while data from~\cite{sasaqui2005} may be attributed to \textit{s}-wave transitions which result in more or less constant rates~\cite{wiescher1990}.

The present data in $\mathrm{cm}^{3}/(\mathrm{s}\cdot\mathrm{mole})$ were parametrized as~\cite{rauscher2000}
\begin{align}
  N_{A} \langle \sigma_{\mathrm{n}\gamma}^{*} v \rangle  = \; & \mathrm{exp} (a_{0} + a_{1}T^{-1}_{9} + a_{2}T^{-1/3}_{9} + a_{3}T^{1/3}_{9} \nonumber \\
                                             & + a_{4}T_{9} + a_{5}T^{5/3}_{9} + a_{6}\mathrm{ln}T_{9}) , \label{eq:para_l}
\end{align}
in the temperature range of interest with $N_{\mathrm{A}}$ being the Avogadro constant. The best fit para\-me\-ters are $a_{0} = 1.019 \cdot 10^{1}, a_{1} = -2.229 \cdot 10^{-2}, a_{2} = 2.849, a_{3} = -6.089, a_{4} = 3.146 \cdot 10^{-1}, a_{5} = -1.564 \cdot 10^{-2}$ and $a_{6} = 3.492$.

In the lower panel of Figure~\ref{fig:rr}, the relative contribution of capture to the ground state in $^{18}$C with respect to the total reaction rate is displayed. This indicates the fraction of the reaction rate that is constrained by the present Coulomb dissociation experiment. Overall, measured data have minor influence. The slight increase towards higher temperature starting at $T_{9}=1$ is due to transitions from excited core states where experimental data get more important than the theoretical contribution of excited states in $^{18}$C.

The experimental spectroscopic factors $S_{i}(J^{\pi}_{\rm core}, l_{j})$ in Table~\ref{tab:xs_c2s} agree with the shell model calculations within a $3\sigma{}$-range and the calculations were employed in the present thermonuclear reaction rates. The parametrization of the reaction rates, see Equation~(\ref{eq:para_l}), with transitions to the ground state of $^{18}$C, which are scaled to the experimentally derived spectroscopic strengths, differ by $10\%$ from the given $a_{i};\;{}i=1\dots{}6$ using spectroscopic amplitudes exclusively from the shell model calculations. This is mainly due to the relatively strong deviation of the amplitudes in the second excited state of $^{17}$C, which becomes more important at higher temperature.


\section{Implications on the \textit{r}-process}
\label{sub:imp}

To explore the impact of the newly derived neutron capture rate of $^{17}$C on the \textit{r}-process nucleosynthesis comprehensive network calculations for parametrized models that may represent different possible astrophysical conditions were performed. The parametrization for the temperature evolution is given by
\begin{equation}
T(t)=
\begin{cases}
T_a+T_0\exp[-(t-t_0)/t_{\rm dyn}], & t \leq t_1\\
T(t_1)\times(t_T-t_2)/(t-t_2), & t>t_1,
\end{cases}
\label{eq_para_t}
\end{equation}
 similar to the one used in \cite{otsuki2002, sasaqui2005}, where $T_0+T_a$ is the initial temperature, and $t_{\rm dyn}$ characterizes the dynamical timescale of the ejecta. The density evolution is derived by assuming a constant radiation-dominated entropy per nucleon
\begin{equation}
s=\frac{11}{45}\frac{\pi^2}{\rho/m_u}\left(\frac{k_B T}{\hbar c}\right)^3.
\end{equation}

\begin{figure}[hbt]
  \centering
   \includegraphics[angle=0,width=1.\columnwidth]{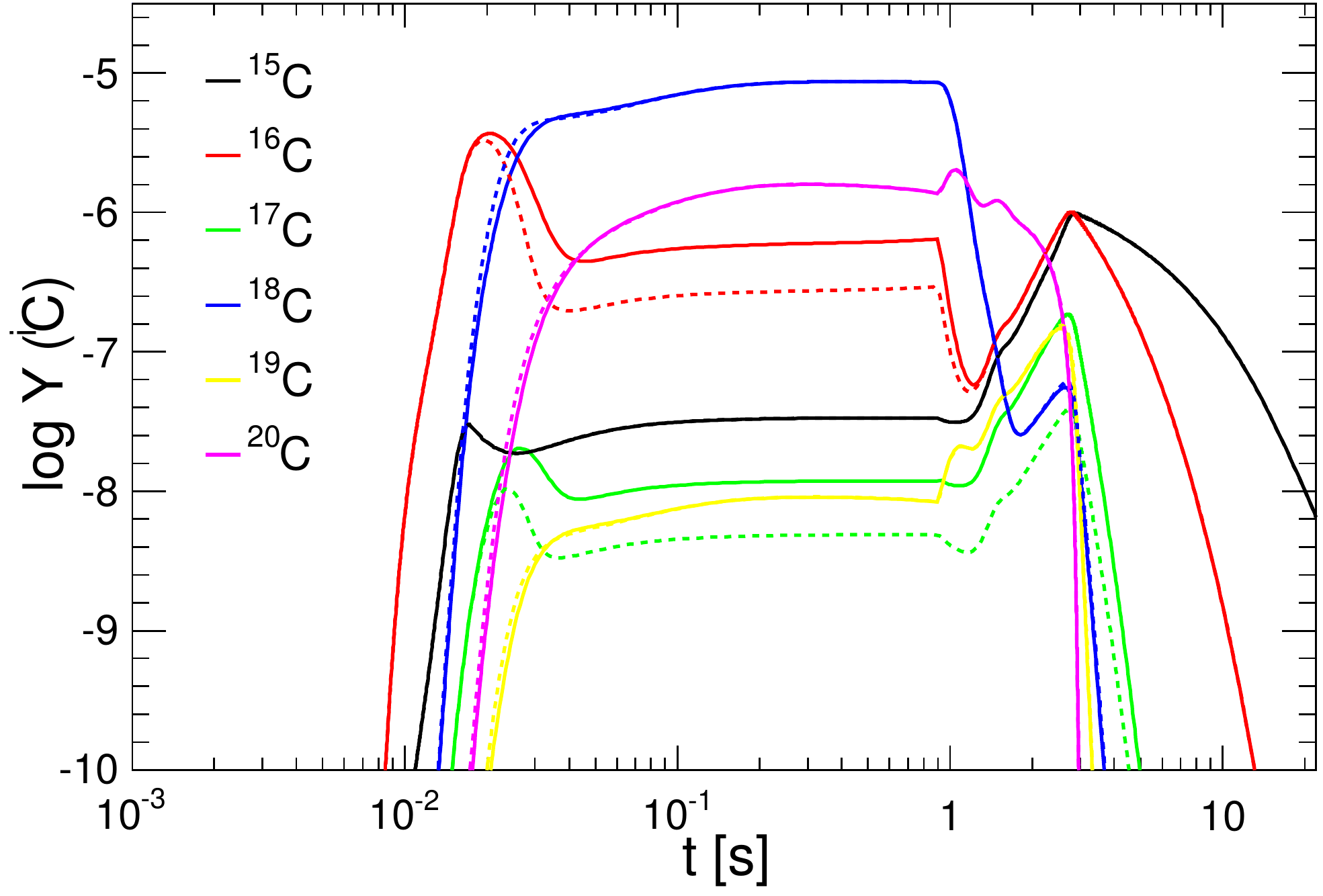}
   \caption{[color online] Evolution of the abundances for carbon isotopes assuming parametric conditions as discussed in the text. Results using the experimentally based $^{17}$C$(\mathrm{n},\gamma)$ (full lines) are compared with the rate from~\cite{sasaqui2005} (dashed lines).
  \label{fig:rate_wu}}
\end{figure}

Here, $t_{\rm dyn}=5$~ms, $s=350$, $Y_e=0.45$, $T_0=8.4$~GK, $T_a=0.6$~GK, $t_0=t_2=0$, and $t_1=1$~s were chosen to represent the high entropy and fast expanding ejecta. That was regarded as the main \textit{r}-process site in the neutrino-driven wind of the core-collapse supernovae~\cite{woosley1994}. Although such an environment has not been achieved in recent simulations~\cite{huedepohl2010, pinedo2014}, it was used in the previous sensitivity study of \cite{sasaqui2005} and is considered here for the purpose of illustration. In Figure~\ref{fig:rate_wu}
the evolution of the abundances of carbon isotopes $Y(^i$C) during the \textit{r}-process is shown based on the rate of $^{17}\mathrm{C(n},\gamma)$ determined in this work (full lines) and the theoretical value used in~\cite{sasaqui2005} (dashed lines). For the conditions of the present parametrization the temperature remains nearly constant around 0.6~GK for times between 0.03~s and 1~s. During this period the abundances of carbon isotopes do not change substantially. The temperature is not large enough to maintain a full $(\mathrm{n},\gamma) \leftrightarrows (\gamma,\mathrm{n})$ equilibrium along the carbon isotopic chain due to the large neutron separation energies of the even $N$ isotopes. Nevertheless, a quasi-equilibrium develops in which the $(\mathrm{n},\gamma)$ and $(\gamma,\mathrm{n})$ reactions connecting an isotope with an even neutron number with the heavier odd neutron number one are in equilibrium. The odd neutron number isotope is connected to the heavier even isotope only by a $(\mathrm{n},\gamma)$ reaction. Under this conditions, changing the $^{17}$C$(\mathrm{n},\gamma)$ reaction affects basically only the abundances of $^{16}$C and $^{17}$C. 

\begin{figure}[hbt]
  \centering
  \includegraphics[angle=0,width=1.\columnwidth]{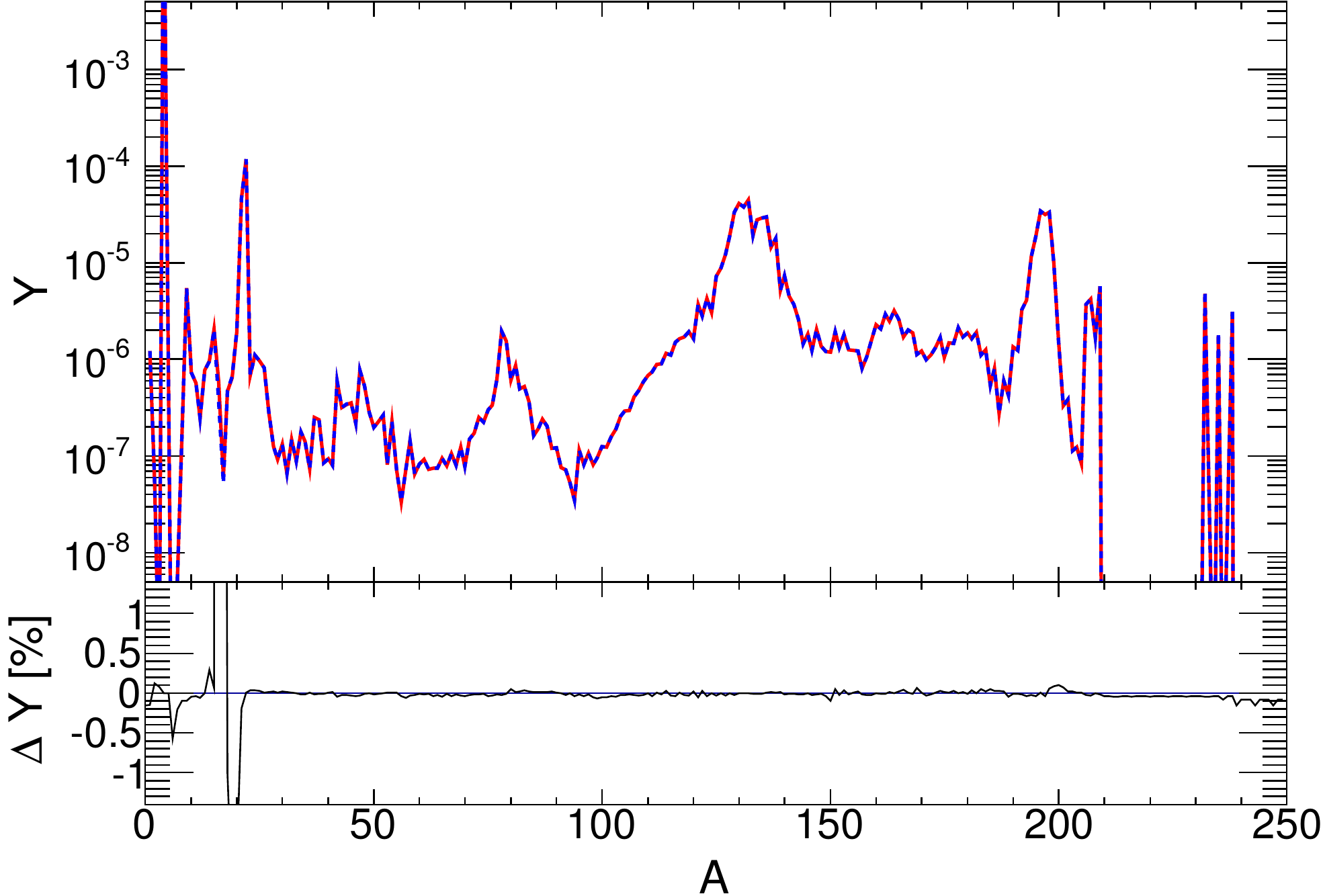}
  \caption{[color online] (Upper panel) Abundance as a function of mass number~\textit{A} at $t\sim 10^{9}$~years with updated $^{17}$C capture rates (red) in comparison to rates from~\cite{sasaqui2005} (blue). In the lower panel the percentage change of abundances calculated with the rate of this work relative to the abundances calculated with the rate from~\cite{sasaqui2005}.
  \label{fig:mass_wu}}
\end{figure}

In Figure~\ref{fig:mass_wu},
the abundances at the end of the \textit{r}-process based on network calculations with two different reaction rates for the $^{17}$C$(\mathrm{n},\gamma)$ are compared: the rate determined in this work and the rate listed in~\cite{sasaqui2005}. It can be seen that although the two rates differs by about a factor of two at $T\sim 0.6$~GK (see Figure~\ref{fig:rr}), the resulting percentage change for the abundances $\Delta Y$ calculated with the rate of this work relative to the abundances calculated with the rate from~\cite{sasaqui2005} is generally less than 0.1\% except for nuclei with $A \lesssim 20$ (lower panel). This can be understood because the build up of nuclei heavier than carbon is governed by the average beta-decay rate $\lambda_{\beta}(\mathrm{C}) = \sum_{A} \lambda_{\beta}(^{A}\mathrm{C}) Y(^{A}\mathrm{C})$ of the carbon isotopic chain. This rate is determined mainly by $^{18}$C and $^{20}$C whose abundances are unchanged (see Figure~\ref{fig:rate_wu}). 

The impact of the two rates for another set of parameters with $t_{\rm dyn}=80$~ms, $s=25$, $Y_e=0.235$, $T_0=12.0$~GK, $T_a=0.25$~GK, $t_0=0.325$~s, $t_1=0.8$~s, and $t_2=0.4$~s was also explored. This condition mimics the matter ejected from the accretion disk formed during binary neutron star mergers \cite{Fernandez2013, just2014}. Such a scenario may contribute significantly to the \textit{r}-process inventory. We again find a similar quasi-equilibrium and the final \textit{r}-process abundances are not sensitive to the change of the neutron capture rate on $^{17}$C.

\section{Summary}
\label{sec:sum}

In the present work the neutron-capture cross section of $^{17}$C was obtained from an indirect measurement complemented by theoretical calculations. The stellar reaction rate was derived accounting for the stellar enhancement of the target nucleus $^{17}$C as well as the feeding of all bound states of $^{18}$C.

Experimental exclusive Coulomb dissociation transitions to excited states in $^{17}$C were identified using prompt gamma-rays in the Crystal Ball detector and ground state transitions were tagged by the absence of the gamma trigger. The differential cross sections were supplemented by theoretical calculations in a semi-classical model for Coulomb dissociation cross sections utilizing a core valence-neutron model. These calculations were used to acquire the complete set of continuum transitions between all bound states in $^{17}$C and $^{18}$C. The calculated energy-differential capture cross sections were weighted with the appropriate Boltzmann factor to obtain the astrophysical reaction rate. The obtained rate differs significantly from recently used parametrizations. The present rate is taken as input to comprehensive \textit{r}-process network calculations.

The implications on the \textit{r}-process were studied assuming conditions corresponding to a high entropy neutrino-driven wind of core-collapse supernovae as well as ejecta from accretion disks formed during the binary neutron star mergers. No notable influence of the neutron-capture rates of $^{17}$C on the final \textit{r}-process abundances has been observed. For the thermodynamical conditions considered, the build up of nuclei heavier than carbon is governed by $^{18}$C or $^{20}$C with large beta-decay rates. Their abundances remain almost unchanged when using the new rate.

\section{Acknowledgement}
\label{sec:akn}

This work was supported by the Helmholtz International Center for FAIR within the framework of the LOEWE program launched by the state of Hesse; by the Helmholtz Alliance Program of the Helmholtz Association, Contract No. HA216/EMMI ???Extremes of Density and Temperature: Cosmic Matter in the Laboratory???; by the GSI-TU Darmstadt Cooperation agreement; by the BMBF under Contracts No. 06DA70471, No. 06DA9040I, and No. 06MT238; by the Helmholtz Association through the Nuclear Astrophysics Virtual Institute, No. VH-VI-417, by the DFG cluster of excellence Origin and Structure of the Universe; by US DOE Grants No. DE-FC02-07ER41457 and No. DE-FG02-96ER40963, via the GSI-RuG/KVI collaboration agreement; by the Portuguese FCT, Project No. PTDC/FIS/103902/2008, the Spanish FPA2012-32443, and by the U.S. NSF Grant 1415656 and the U.S. DOE Grant DE-FG02-08ER41533.

\bibliographystyle{prc-format}
\bibliography{Bib/prc_bib_paper}

\end{document}